\documentclass[journal abbreviation]{copernicus}

\usepackage{color}
\usepackage[T1]{fontenc}

\newcommand{\MC}{\mathcal}

\begin{document}

\title{Causal Kinetic Equation of Non-equilibrium Plasmas%by seed-field generation
}

\author[1,2]{R. A. Treumann}
%\author[3]{R. Nakamura}
\author[3]{W. Baumjohann}
%\author[2]{Y. Narita}

\affil[1]{Department of Geophysics and Environmental Sciences, Munich University, Munich, Germany}
\affil[2]{International Space Science Institute Bern, Switzerland}
%% The [] brackets identify the author to the corresponding affiliation, 1, 2, 3, etc. should be inserted.
\affil[3]{Space Research Institute, Austrian Academy of Sciences, Graz, Austria}

\runningtitle{Causal Kinetic Theory}

\runningauthor{R. A. Treumann and W. Baumjohann}

\correspondence{R. A.Treumann\\ (art@geophysik.uni-muenchen.de)}

\received{ }
%\pubdiscuss{ } %% only important for two-stage journals
\revised{ }
\accepted{ }
\published{ }

%% These dates will be inserted by the Publication Production Office during the typesetting process.

\firstpage{1}

\maketitle

%\begin{abstract}
{\bf Abstract. --} Statistical plasma theory far from thermal equilibrium is subject to Liouville's equation which is at the base of the BBGKY hierarchical approach to plasma kinetic theory from which in the absence of collisions Vlasov's equation follows. It is also at the base of Klimontovich's approach which includes single particle effects like spontaneous emission. All these theories have been applied to plasmas with admirable success even though they suffer from a fundamental omission in their use of the electrodynamic equations in the description of the highly dynamic interactions in many-particle particle conglomerations. In the following we extend this theory to  taking into account that the interaction between particles separated from each other at a distance requires the transport of information. Action needs to be transported and thus in the spirit of the direct-interaction theory as developed by \citet{wheeler1945} requires time. This is done by reference to the retarded potentials. We derive the fundamental causal Liouville equation for the phase space density of a system composed of a very large number of charged particles. Applying the approach of \citet{klimontovich1967} we obtain the retarded time evolution equation of the one-particle distribution function in plasmas which replaces Klimontovich's equation in cases when the direct-interaction effects have to be taken into account. This becomes important in all systems where the distance between two points $|\Delta \vec{q} | \sim ct$ is comparable to the product of observation time and light velocity, which is typical in cosmic physics and astrophysics.
% \keywords{MHD turbulence, turbulent dispersion relation, turbulent response function, solar wind turbulence}
%\end{abstract}

%\vspace{0.5cm}
\section{Introduction}
The starting point of (classical) kinetic theory is Liouville's equation. Written in terms of the $N_a$-particle Hamiltonian $\MC{H}_{N_a}(\vec{q,p},t)$ and defining the 6d-phase space density $\MC{N}_a(\vec{q,p,}t)$ of species $a$, both functions of space $\vec{q}$ and momentum $\vec{p}$, it becomes
\begin{equation}\label{eq-1}
{\dot\MC{N}}_a\equiv\partial_t\MC{N}_a+\big[\MC{H}_{N_a},\MC{N}_a\big]=0
\end{equation}
where it is assumed that the particle number $N_a$ of species $a$ is conserved (along all their dynamical phase-space orbits). Otherwise the right-hand side would contain the difference of number sources and losses $\MC{S}_a-\MC{L}_a$. This equation, under the assumptions made, is completely general applying to any system consisting of $N_a\gg1$ particles in interaction with an external as well as with their mutual fields of which they function as sources. These fields are contained in the Hamiltonian and act via the Poisson bracket $[\dots]$. 

In view of application to plasmas the relevant field is the electromagnetic field $\vec{E,B}$ with the particles carrying electric charges $e_a=\mp e\equiv ae$ (with $a=-,+$) being the sources of the field. For simplicity we, in the following, restrict to electrons and ions (protons) of respective mass $m_a$, and gravity can be neglected on all scales small enough for the electromagnetic fields to dominate. We also assume global quasineutrality and absence of any external fields. Then $\vec{E,B}=(\vec{E,B})^m$ is the set of \emph{microscopic} electromagnetic fields produced solely by the microscopic charge and current densities of the interacting particle components $a$ which serve as their sources
\begin{eqnarray}\label{eq-2}
\rho_a^m(\vec{q},t)&=&e_a\sum_{i=1}^{N_a}\delta\big(\vec{q}-\vec{q}_{ai}(t)\big)=e_a\int d^3p\ \MC{N}_a(\vec{p,q},t) \nonumber\\[-0.5ex]
\vec{j}^m_a(\vec{q},t)&=&\frac{e_a}{m_a}\sum_{i=1}^{N_a}\ \vec{p}_{ai}(t)\ \delta\big(\vec{q}-\vec{q}_{ai}(t)\big)= \\[-0.25ex]
&=&\frac{e_a}{m_a}\int d^3p\ \vec{p}\ \MC{N}_a(\vec{p,q},t) \nonumber
\end{eqnarray}
where the exact 6d-phase space density is defined through 
\begin{equation}\label{eq-3}
\MC{N}_a(\vec{p,q},t)=\sum_{i=1}^{N_a}\delta\big(\vec{p}-\vec{p}_{ai}(t)\big)\ \delta\big(\vec{q}-\vec{q}_{ai}(t)\big)
\end{equation}
and $\vec{q}_{ai}(t),\vec{p}_{ai}(t)$ are the spatial and momentum phase-space trajectories which the particle $ai$ performs in the phase space under the action of the complete microscopic electromagnetic field $(\vec{E,B})^m$ which it feels at its location $\vec{q}=\vec{q}_{ai}(t)$ at time $t$. 
Liouville's equation for the exact phase-space density can then be written in the form
\begin{eqnarray}\label{eq-4}
\frac{\partial\MC{N}_a}{\partial t}+\frac{\vec{p}}{m_a}\vec{\cdot}\nabla_{\vec{q}}\ \MC{N}_a&\!\!\!\!+\!\!\!\!&e_a\bigg\{\Big[\vec{E}^m(\vec{q},t)\nonumber\\[-1ex]
&&\\[-1ex]
&+&\frac{\vec{p}}{m_a}\vec{\wedge}\vec{B}^m(\vec{q},t)\Big]\vec{\cdot}\: \frac{\partial\MC{N}_a}{\partial\vec{p}}\bigg\}=0\nonumber
\end{eqnarray}
This is Klimontovich's equation for the exact microscopic phase space density $\MC{N}_a(\vec{p,q},t)$ in 6d-phase space \citep{klimontovich1967}. It is a tautology because it does not say anything other than that particle number is conserved along all the dynamical orbits of the particles in phase space under the action of their mutual electromagnetic fields. The microscopic fields it contains are given by Maxwell's equations in differential form
\begin{eqnarray}
\nabla_q\vec{\wedge}\:\vec{B}^m&=&\mu_0\vec{j}^m+\mu_0\epsilon_0\partial_t\vec{E}^m,\quad \nabla_q\vec{\cdot}\:\vec{B}^m=0\nonumber\\[-1.5ex]
&&\\[-1.5ex]
\nabla_q\vec{\wedge}\:\vec{E}^m&=&-\partial_t\vec{B}^m,\quad \nabla_q\vec{\cdot}\:\vec{E}^m=\frac{1}{\epsilon_0}\sum_a\rho_a^m\nonumber
\end{eqnarray}
Solution of this set of equations is not possible as it requires knowledge of all microscopic particle orbits. One can, however introduce some coarse graining procedure and define integrated distribution functions which ultimately reduce the system to a set of equations known as Klimontovich-Vlasov equations for a one particle phase space distribution in the presence of the average electromagnetic fields. This procedure is very efficient, and we will follow it below in a modified version. 

\section{Effect of retardation}
The problem of the above equations is that it does not account for the fact that the electromagnetic signal of presence and motion of the particles is transferred from the signal-emitting particles to the signal receiving particles under consideration, the absorbers and reactors. Their sources are the charge and current densities $\rho^m_a(\vec{q},t),\: \vec{j}^m(\vec{q},t)$ which are assumed to be known at any instant $t$ in all space points $\vec{q}$. Obtaining this knowledge is impossible as it requires instantaneous measurements at time $t$ of all positions $\vec{q}$ and momenta $\vec{p}$ of the particles present in real space. Instead, the information must be synchronized among all locations. This is taken care of in the Li\'enard-Wiechert potentials which explicitly account for the transport of information from point $\vec{q}'$ to point $\vec{q}$. In this case in the Lorentz gauge
\begin{eqnarray}\label{eq-6}
\vec{E}^m&=& -\nabla_{\vec{q}}\phi^m(\vec{q},t)-\partial_t\vec{A}^m(\vec{q},t)\nonumber\\[-1.5ex]
&&\\[-1.5ex] 
\vec{B}^m&=&\nabla_{\vec{q}}\vec{\wedge}\vec{A}^m(\vec{q},t)\nonumber
\end{eqnarray}
the correct scalar and vector potentials are to be expressed by the \emph{retarded} charge and current densities 
\begin{eqnarray}\label{eq-7}
\phi^m(\vec{q},t)&=&\frac{1}{4\pi\epsilon_0}\sum_a\int d^3q'\ \frac{\rho^m_a(\vec{q}',t')}{|\vec{q}-\vec{q}'|} \nonumber \\[-1.5ex]
&&\\[-1.5ex]\nonumber
\vec{A}^m(\vec{q},t)&=&\frac{\mu_0}{4\pi}\sum_a\int d^3q'\ \frac{\vec{j}^m_a(\vec{q}',t')}{|\vec{q}-\vec{q}'|}\nonumber
\end{eqnarray}
taken at the retarded time 
\begin{equation} \label{eq-8}
t'=t-|\vec{q}-\vec{q}'|/c, \qquad \vec{q}\neq\vec{q}'
\end{equation}
of arrival of \emph{all the signals} emitted at $t'$ from all the particles at spatial distance $|\vec{q}-\vec{q}'|$ from  the location of particle $a_i$ at $\vec{q}$ and at time $t$. This also implies that in the expressions for the charge and current densities $\MC{N}_a\to\MC{N}_a(\vec{p},\vec{q}',t')$ is a function of the retarded time $t'$.

Since all particles serve both as field sources and actors, excluding their self-interaction, the use of the instantaneous fields ignores the time-consuming signal transport and thus cannot be correct. It is an approximation only that holds for comparably small volumes such that, in the expression for the retarded time, the spatial difference can be neglected. Thus the restriction on the distance between particles is that 
\begin{equation}
|\Delta q|\ll c\Delta t 
\end{equation}
Clearly, this condition will readily be violated in large volumes of cosmic and astrophysical size, where one must refer to the above precise potentials and the fields resulting from them in reference to the Lorentz gauge.

This problem has for single particle-particle interactions be discussed in depth in seminal papers by \citet{wheeler1945,wheeler1949}. They showed that in a closed system where no information is lost to the outside eliminating any (ridiculous) self-interaction of a particle with its proper electromagnetic field implies that the fields are properly described via \emph{retarded} potentials only as done above. These account for the emission of a signal by one particle and the absorption of the signal after some travel time by the target particle causing this particle to interact. The emitted signals belong to advanced potentials which when correctly included subtract out thereby restauring the required real world causality. It is incorrect to assume that the information arrives microscopically instantaneously at the target causing this to act. The electromagnetic fields following from the Lorentz gauge in the microscopic domain are
\begin{eqnarray}\label{eq-10}
\vec{E}^m(\vec{q},t)&=&\frac{1}{4\pi\epsilon_0}\int d^3q'\bigg[\bigg(\frac{\rho^m(\vec{q}',t')}{|\vec{q}-\vec{q}'|^3}+\nonumber\\
&+&\frac{\partial_t\,\rho^m(\vec{q}',t')}{c|\vec{q}-\vec{q}'|^2}\bigg)\ (\vec{q}-\vec{q}')-\frac{\partial_t\, \vec{j}^m(\vec{q}',t')}{c^2|\vec{q}-\vec{q}'|}\bigg]\\
\vec{B}^m(\vec{q},t)&=&\frac{\mu_0}{4\pi}\int d^3q'\bigg[\frac{\vec{j}^m(\vec{q}',t')}{|\vec{q}-\vec{q}'|^3}+\frac{\partial_t\, \vec{j}^m(\vec{q}',t')}{c|\vec{q}-\vec{q}'|^2}\bigg]\vec{\wedge}\big(\vec{q}-\vec{q}'\big)\nonumber
\end{eqnarray}
which were independently given first by \citet{panofsky1962} and \citet{jefimenko1966}. One should note that in these expressions the charge $\rho^m$ and current densities $\vec{j}^m$ are summed over all particle species $a$. 

This explicit representation of the microscopic fields accounts properly for the time delay between the signal emitted from the total compound of primed particles to arrive at the location $\vec{q}$ of the particle under consideration. Since the microscopic charge and current densities are functionals of the phase space density these expressions contain the latter, though in a more involved manner than when using the differential forms of the electrodynamic equations which do not show where and whether the retardation of the signal is taken into account. It is clear from these expressions that particles which are far away from the target do not affect it. The main effect will always come from close neighbours. 

\section{Retarded charge and current densities}
Taking the divergence of the microscopic electric field and the curl of the microscopic magnetic field one readily reads the correct microscopic charge and current densities when comparing the expressions with the microscopic Maxwell equations:
\begin{eqnarray}\label{eq-11}
\rho^m(\vec{q},t)&=&\frac{1}{4\pi}\sum_a\nabla_q\cdot\int d^3q'\ \bigg[\bigg(\frac{\rho^m_a(\vec{q}',t')}{|\vec{q}-\vec{q}'|^3}+\nonumber\\[-1.5ex]
&&\\[-1.5ex]
&&+\ \frac{\partial_t\,\rho^m_a(\vec{q}',t')}{c|\vec{q}-\vec{q}'|^2}\bigg)\ \big(\vec{q}-\vec{q}'\big)-\frac{\partial_t\,\vec{j}^m_a(\vec{q}',t')}{c^2|\vec{q}-\vec{q}'|}\bigg]\nonumber\\
\vec{j}^m(\vec{q},t)&=&\frac{1}{4\pi}\sum_a\bigg\{\nabla_q\vec{\wedge}\int d^3q'\bigg[\frac{\vec{j}^m_a(\vec{q}',t')}{|\vec{q}-\vec{q}'|^3}+\frac{\partial_t\,\vec{j}^m_a(\vec{q}',t')}{c|\vec{q}-\vec{q}'|^2}\bigg]\nonumber\\
&&\vec{\wedge}\ \big(\vec{q}-\vec{q}'\big)-\partial_t\int d^3q'\ \bigg[\bigg(\frac{\rho^m_a(\vec{q}',t')}{|\vec{q}-\vec{q}'|^3}+\\
&&+\frac{\partial_t\,\rho^m_a(\vec{q}',t')}{c|\vec{q}-\vec{q}'|^2}\bigg)\ \big(\vec{q}-\vec{q}'\big)-\frac{\partial_t\, \vec{j}^m_a(\vec{q}',t')}{c^2|\vec{q}-\vec{q}'|}\bigg]\bigg\}\nonumber
\end{eqnarray}

These are the correct forms of the summed over species $a$ charge and current densities which have to be used in Maxwell's equations in order to account for the retarded transfer of information between the particles in the plasma. These expressions are implicit for both the charge and current densities. In order to relate them to the exact microscopic phase space distribution $\MC{N}_a$ as defined in Eq. (\ref{eq-3}) one refers to the representations (\ref{eq-2}) of the charge and current densities. This shows that the functional dependence of the phase-space density is itself implicit. It depends on itself taken at all the positions $\vec{q}'$ and retarded times $t'$.  

The proper way of dealing with this problem is to stay as long as possible in the microscopic picture. There all the charged particles can be imagined as moving in vacuum, if only the medium is sufficiently dilute. By progressing to a coarse-grained picture one may afterwards step up to considering a more continuous medium in which ultimately the propagation properties of the signals will become modified by the collective properties of the matter. 

With these results it is convenient to express the microscopic electromagnetic fields through the microscopic phase space densities of the particle species 
\begin{eqnarray}\label{eq-13}
\vec{E}^m(\vec{q},t)&=&\sum_a\frac{e_a}{4\pi\epsilon_0}\int d^3p\ d^3q'\bigg[\bigg(\frac{\MC{N}_a(\vec{p,q}',t')}{|\vec{q}-\vec{q}'|^3}+\nonumber\\[-2.5ex]
&&\\
&+&\frac{\partial_t\,\MC{N}_a(\vec{p,q}',t')}{c|\vec{q}-\vec{q}'|^2}\bigg)\ (\vec{q}-\vec{q}')-\frac{\vec{p}\ \partial_t\,\MC{N}_a(\vec{p,q}',t')}{c^2|\vec{q}-\vec{q}'|}\bigg]\nonumber\\
\vec{B}^m(\vec{q},t)&=&\sum_a\frac{e_a\mu_0}{4\pi}\int d^3p\ d^3q'\bigg[\frac{\vec{p}\MC{N}_a(\vec{p,q}',t')}{|\vec{q}-\vec{q}'|^3}+\nonumber\\[-1.5ex]
&&\\[-1.5ex]
&+&\frac{\vec{p}\ \partial_t\,\MC{N}_a(\vec{p,q}',t')}{c|\vec{q}-\vec{q}'|^2}\bigg]\vec{\wedge}\big(\vec{q}-\vec{q}'\big)\nonumber
\end{eqnarray}
These are the expressions of the electromagnetic field which have to be used in the microscopic Liouville equation (\ref{eq-4}) for the microscopic $N$-particle phase space density. Not only that they couple the different particle species thus leading to a coupling between their phase space distributions, they also make each microscopic distribution $\MC{N}_a$ a functional of the distributions taken at all different phase space locations which are causally accessible via their retarded times of signal propagation $t'$. Clearly this is a substantial complication which is introduced into kinetic theory by the requirement of causality.

It is quite inconvenient to deal with all microscopic phase space densities. We would rather have an equation for they of the separately. This can be achieved when observing that Eq. (\ref{eq-4}) is an equation for $\MC{N}_a$. Thus putting $a\to b$ in the last expressions which means that we sum over all particle species $b$ including also $b=a$ (with self-interaction excluded by the definition of the retarded time) we have
\begin{eqnarray}\label{eq-15}
\,\frac{\partial\:\MC{N}_a}{\partial t}&+&\frac{\vec{p}}{m_a}\vec{\cdot}\nabla_{\vec{q}}\:\MC{N}_a+\sum_b\frac{e_ae_b}{4\pi}\int d^3p'\ d^3q'\Bigg\{\nonumber\\
&&\frac{1}{\epsilon_0}\bigg[\bigg(\frac{\MC{N}_b(\vec{p',q}',t')}{|\vec{q}-\vec{q}'|^3}
+\frac{\partial_t\,\MC{N}_b(\vec{p',q}',t')}{c|\vec{q}-\vec{q}'|^2}\bigg)\ \big(\vec{q}-\vec{q}'\big)-\nonumber\\
&-&\frac{\vec{p'}\ \partial_t\,\MC{N}_b(\vec{p',q}',t')}{c^2|\vec{q}-\vec{q}'|}\bigg]
-\frac{\mu_0}{m_a}\vec{p}\vec{\wedge}\bigg[\big(\vec{q}-\vec{q}'\big)\vec{\wedge}\\
&&\bigg(\frac{\vec{p'}\MC{N}_b(\vec{p',q}',t')}{|\vec{q}-\vec{q}'|^3}+ \frac{\vec{p'}\ \partial_t\,\MC{N}_b(\vec{p',q}',t')}{c|\vec{q}-\vec{q}'|^2}\bigg)\nonumber\\
&&\bigg]\Bigg\}\:{\vec\cdot}\:\frac{\partial}{\partial\vec{p}}\MC{N}_a\big(\vec{p,q},t\big)=0\nonumber
\end{eqnarray}
Here $\MC{N}_a(\vec{p,q},t)$, while $\MC{N}_b(\vec{p}',\vec{q}',t')$ depends on the dummy coordinates of all particles $b$ of integration and on the retarded time $t'=t-|\vec{q}-\vec{q}'|/c,\ \vec{q}\neq\vec{q}'$. Thus in the $q'$ integration also all particles of sort $a$ are included with the exception of the particle located at $\vec{q}=\vec{q}'$ at time $t$.

The above equation (\ref{eq-15}) is the \emph{causal} Liouville equation acting on the microscopic $N_a$-particle phase space density $\MC{N}_a(\vec{p,q},t)$ in the presence of a large number of charged particles interacting via their self-consistently generated electromagnetic fields. It extends Klimontovich's equation to the correct inclusion of the retardation effect of transmission of information between the particles via electromagnetic fields. 

Inclusion of information transport between the interacting particles substantially complicates the basic kinetic equation. It causes delay of response and thus refers to a natural measuring process in which the particles are not only generators of the electromagnetic field but also measure its effect over a causal distances accessible to them. The delay must thus necessarily cause decorrelation of the response.

There is another complication with this picture which comes into play when considering large compounds of particles rather than single particles. Single charged  particles are assumed to move in the vacuum, the signal propagation between them takes place at light speed $c$. Immersed into a comparably dense environment of all the other charged particles any light respectively radiation experiences radiation transport which is dominated by scattering, reflection, transmission and absorption, processes that occur due to the active response of the environment to the presence of radiation and depend on the capabilities of the medium to let electromagnetic signals pass. In these processes various proper electromagnetic modes excited in the medium become involved. These are solutions of the dispersion properties of the matter. Hence correctly accounting for the signal transport becomes rather involved. For this reason the theory even in this complex version applies to sufficiently dilute media to allow the assumption of signal propagation in vacuum.

In the following we will proceed along the same lines as \citet{klimontovich1967} but will in the end refer to the above field equations. This means that in defining the average distributions we will consider Liouvilles equation without explicit reference to the fields.

\section{Average distribution functions}
Dealing with the causal $N$-particle kinetic equation (\ref{eq-15}) is impractical. One wants to reduce it to an equation for a one-particle distribution function in 6d-phase space for indistinguishable particles of sort $a$. This is done by integrating out in Eq. (\ref{eq-3}) all particle coordinates $i>1$. Defining phase space coordinates $\vec{x}=(\vec{p},\vec{q}),\ \vec{x}_{ai}(t)=\big(\vec{p}_{ai}(t),\:\vec{q}_{ai}(t)\big)$ the $N$-particle density becomes
\begin{equation}\label{eq-16}
\MC{N}_a(\vec{x}, t)=\sum_{i=1}^{N_a}\delta\big(\vec{x}-\vec{x}_{ai}(t)\big)
\end{equation}
Following \citet{klimontovich1967} let us define the one-particle distribution of sort $a$ of indistinguishable particles by
\begin{eqnarray}\label{eq-17}
f_a(\vec{x}_{a1}, t) &=&V_a\int f_Nd^6\vec{x}_{a2}\dots d^6\vec{x}^6_{aN_a}\times \nonumber\\[-1ex]
&&\\[-1ex]
&&\times\prod_{b\neq a}d^6\vec{x}_{b1}\dots d^6\vec{x}^6_{bN_b}\nonumber
\end{eqnarray}
The $N$-particle probability distribution $f_N$ depends on all the particle coordinates in phase space which have been integrated out in the last expression including $\vec{x}_{a1}$, and $V$ is the spatial volume of particle $a1$, i.e. the volume all indistinguishable particle occupy. With its help the averaged phase space density yields directly
\begin{equation}\label{eq-18}
\frac{N_a}{V_a}f_a(\vec{p},\vec{q},t)=\big\langle\MC{N}_a(\vec{p,q},t)\big\rangle
\end{equation}
Here the right-hand side is the ensemble-averaged one-particle phase-space density $\langle\MC{N}_a(\vec{p},\vec{q},t)\rangle$ which is a functional only of the indistinguishable dynamics of the particles indexed by $i=1$. Accordingly, averaging the product of two phase space densities $\MC{N}_a(\vec{x},t)$ and $\MC{N}_b(\vec{x}',t)$ yields
\begin{eqnarray}\label{eq-19}
\big\langle\MC{N}_a(\vec{x},t)\MC{N}_b(\vec{x}',t)\big\rangle &=& n_a\delta_{ab}\delta(\vec{x}-\vec{x}')f_a(\vec{x},t)+\nonumber\\ &+&n_an_bf_{ab}(\vec{x,x}',t)
\end{eqnarray}
where the partial densities are defined as $n_a=N_a/V_a,\ n_b=N_b/V_b$, and $f_{ab}(\vec{x},\vec{x}',t)$ is the two-particle distribution function. In the same way higher order average products of phase space densities can be reduced to sums of distribution functions.

This procedure must be applied to the causal $N$-particle kinetic equation (\ref{eq-15}). This is a formidable task if using the $N$-particle kinetic equation in its explicit form. As announced earlier it is more convenient to remain with the implicit versions of the Lorentz gauge (\ref{eq-6}) and the retarded potentials in which we replace the charge end current densities by the general expressions given in Eq. (\ref{eq-2}). This yields from (\ref{eq-7}) 
\begin{eqnarray}\label{eq-20}
\phi^m(\vec{q},t)&=&\sum_a\frac{e_a}{4\pi\epsilon_0}\int d^3p\,d^3q'\:\frac{\MC{N}_a(\vec{p},\vec{q}',t')}{|\vec{q}-\vec{q}'|} \nonumber\\[-1.5ex]
&&\\[-1.5ex]
\vec{A}^m(\vec{q},t)&=&\sum_a\frac{e_a\mu_0}{4\pi}\int d^3p\,d^3q'\:\frac{\vec{p}\ \MC{N}_a(\vec{p},\vec{q}',t')}{|\vec{q}-\vec{q}'|} \nonumber
\end{eqnarray}
with the time taken as the retarded time $t'$ thus depending on the spatial coordinate $\vec{q}'$ which is to be integrated out. 

\section{Causal one-particle kinetic equation}
These expressions are to be used in the Lorentz gauge (\ref{eq-6}) when expressing the electromagnetic fields in the $N$-particle kinetic equation (\ref{eq-4}). Formally this is the same as if we would use (\ref{eq-15}) directly in deriving the corresponding causal equation for the one-particle distribution function of indistinguishable particles of sort $a$. It is only the electromagnetic fields in (\ref{eq-4}) which depend on the retarded time. Therefore one can formally perform the average to obtain
\begin{eqnarray}\label{eq-21}
\frac{\partial\big\langle\MC{N}_a\big\rangle}{\partial t}&+&\frac{\vec{p}}{m_a}\vec{\cdot}\nabla_{\vec{q}}\big\langle\MC{N}_a\big\rangle +\nonumber\\[-1.5ex]
&&\\[-1.5ex]
&+&e_a\bigg\langle\Big[\vec{E}^m(\vec{q},t)+\frac{\vec{p}}{m_a}\vec{\wedge}\vec{B}^m(\vec{q},t)\Big]\vec{\cdot}\frac{\partial\MC{N}_a}{\partial\vec{p}}\bigg\rangle=0\nonumber
\end{eqnarray}
The last term in this equation contains particles of kind $a$ and $b$ as well as the retarded time coordinate. Nevertheless by carefully ordering the different contributions and variables of integration one can bring it into a more convenient form. For this we indicate all integration variables by primes $'$ and rename the retarded variables by a superscript $R$. Then $t'\to t^R=t-|\vec{q}-\vec{q}'|/c$. This yields after expressing the last term in angular brackets for the average phase space density
\begin{eqnarray}\label{eq-22}
\frac{\partial\big\langle\MC{N}_a\big\rangle}{\partial t}&+&\frac{\vec{p}}{m_a}\vec{\cdot}\nabla_{\vec{q}}\big\langle\MC{N}_a\big\rangle\ -\ \sum_b\frac{e_ae_b}{4\pi\epsilon_0}\int d^3p'd^3q'\Bigg\{\nonumber\\[-1ex]
\Big(\nabla_{\vec{q}}&+&\frac{\vec{p}'}{m_bc^2}\frac{\partial}{\partial t}\Big)\ -\ \frac{1}{m_bc^2}\nabla_{\vec{q}} \vec{\wedge}\vec{p}'\ \Bigg\}\ \vec{\cdot}\\
&&\qquad\quad\vec{\cdot}\ \frac{\partial}{\partial\vec{p}}\Bigg(\frac{\big\langle\MC{N}_a(\vec{p},\vec{q},t)\MC{N}_b(\vec{p}',\vec{q}',t^R)\big\rangle}{|\vec{q}-\vec{q}'|}\Bigg)=0\nonumber
\end{eqnarray}
In this version of the phase-space (ensemble) averaged equation for the time and one-particle phase space evolution of the (ensemble) averaged one-particle phase-space density $\langle\MC{N}_a(\vec{p},\, \vec{q},\, t)\rangle$ the retarded time appears only in the averaged product. This equation is the master equation for constructing the kinetic equation for the particle distribution function. Defining the fluctuation of particle number density as $\delta\MC{N}_a(\vec{x},t)=\MC{N}_a(\vec{x},t)-\big\langle\MC{N}_a(\vec{x},t)\big\rangle$ and referring to the correlation function $g_{ab}(\vec{x},\vec{x}',t)$ defined through the average of the product of the fluctuations
\begin{eqnarray}\label{eq-23}
\big\langle\delta\MC{N}_a(\vec{x},t)\:\delta\MC{N}_b(\vec{x}'\!\!,t)\big\rangle&=&n_an_bg_{ab}(\vec{x},\vec{x}'\!\!,t)\ + \nonumber\\
&+&n_a\delta_{ab}\delta(\vec{x}-\vec{x}')f_a
\end{eqnarray}
we finally arrive at the wanted causal kinetic equation
\begin{eqnarray}\label{eq-24}
\frac{\partial f_a}{\partial t}&+&\frac{\vec{p}}{m_a}\vec{\cdot}\nabla_{\vec{q}}f_a-\sum_b\frac{n_be_ae_b}{4\pi\epsilon_0}\int d^3p'd^3q'\Bigg\{\nonumber\\
&&\bigg(\nabla_{\vec{q}}+\frac{\vec{p}'}{m_bc^2}\frac{\partial}{\partial t}\bigg)-\frac{1}{m_bc^2}\nabla_{\vec{q}} \vec{\wedge}\vec{p}'\Bigg\}\ \vec{\cdot}\\
&\vec{\cdot}&\frac{\partial}{\partial\vec{p}}\Bigg(\frac{f_a(\vec{p},\vec{q},t)f_b(\vec{p}',\vec{q}',t^R)}{|\vec{q}-\vec{q}'|}\Bigg)\ =\ \vec{C}_a(\vec{p},\vec{q},t)\nonumber
\end{eqnarray}
The interaction term on the right hand side arises from the various interparticle collisions which are mediated by the electromagnetic field. From the above definition of the fluctuations and correlations it is given by
\begin{eqnarray}\label{eq-25}
\vec{C}_a(\vec{p},\vec{q},t)&\equiv&\ \frac{1}{n_a}\sum_b\frac{n_be_ae_b}{4\pi\epsilon_0}\int d^3p'd^3q'\Bigg\{\nonumber\\[-1ex]
&&\bigg(\nabla_{\vec{q}}+\frac{\vec{p}'}{m_bc^2}\frac{\partial}{\partial t}\bigg)-\frac{1}{m_bc^2}\nabla_{\vec{q}} \vec{\wedge}\vec{p}'\Bigg\}\:\vec{\cdot}\nonumber\\[-1ex]
&&\\[-0.5ex]
&&\vec{\cdot}\frac{\partial}{\partial\vec{p}}\Bigg(\frac{\big\langle\delta\MC{N}_a(\vec{p},\vec{q},t)\ \delta\MC{N}_b(\vec{p}',\vec{q}',t^R)\big\rangle}{|\vec{q}-\vec{q}'|}\Bigg)\nonumber
\end{eqnarray}
Formally these expressions, as claimed in the previous sections, are rather similar to those which, for the non-retarded interactions, had been obtained already by \citet{klimontovich1967} with the only exceptions that here they are written in terms of the full electromagnetic field and contain the spatial integration over all the remote particle space. They, however, are very different from those because they account for the necessary causal relation between the interacting particles which is contained in their dependence on the retarded time $t^R$ by which the particles respond to the transport of information. As a result of this response the spatial integral appearing in these expressions contains an integration over $t^R$. This complicates the calculation substantially and in an analytical treatment requires introduction of further approximations. Nevertheless, the above final equation with the implictly given collision term extends Klimontovich's theory to the explicit reference to causality. 

Referring to Eq. (\ref{eq-22}) the collision term can also be expressed via the fluctuations of the phase space density $\delta\MC{N}_a=\MC{N}_a-\langle\MC{N}_a\rangle$ and the fluctuations of the electromagnetic fields $(\delta\vec{E},\delta\vec{B})=(\vec{E,B})^m-\langle(\vec{E,B})^m\rangle$. This yields
\begin{eqnarray}
\vec{C}_a\ =\ -\frac{e_a}{n_a}\ \int d^3q'd^3p'\!\!\!\!\!\!\!\!&&\!\!\!\!\!\!\!\!\bigg(\frac{\partial}{\partial\vec{p}}\big\langle\delta\vec{E}\delta\MC{N}_a\big\rangle -\nonumber\\
&-&\frac{\vec{p}}{m_a}\ \vec{\cdot}\ \frac{\partial}{\partial\vec{p}}\vec{\wedge}\big\langle\delta\vec{B}\delta\MC{N}_a\big\rangle\bigg)\nonumber
\end{eqnarray}
where the average refers to the integration over all particle space $i>1$, and all quantities still depend on the retarded time $t^R$ which requires integration with respect to $q'$. It is, however, more convenient to make use of the representation via the correlation function, in which case we have from Eqs. (\ref{eq-23}, \ref{eq-25})
\begin{eqnarray}
\vec{C}_a(\vec{p},\vec{q},t)&\equiv&\sum_b\frac{n_be_ae_b}{4\pi\epsilon_0}\int d^3p'd^3q'\Bigg\{\bigg(\nabla_{\vec{q}}+\frac{\vec{p}'}{m_bc^2}\frac{\partial}{\partial t}\bigg)-\nonumber\\[-1ex]
&&\\[-1ex]
&-&\frac{1}{m_bc^2}\nabla_{\vec{q}} \vec{\wedge}\vec{p}'\Bigg\}\ \vec{\cdot}\ \frac{\partial}{\partial\vec{p}}\frac{g_{ab}(\vec{p},\vec{p}',\vec{q},\vec{q}',t^R)}{|\vec{q}-\vec{q}'|}\nonumber
\end{eqnarray}
This is the general causal collision integral term including the interactions between particles indexed by $i=1$ and $i=2$.

From all these expressions one can again obtain an equation for the fluctuation of phase space density $\delta\MC{N}_a$ as well as for the fluctuating fields expressed through the space charge density and current fluctuations. 

The equation (\ref{eq-24}), with knowledge of the collision term on the right or some of its approximations provides the basis for a linearized kinetic theory to any order including particle and time-retarded interaction effects. For this one defines the fluctuations of the one-particle distribution function in the usual way as
\begin{equation}\label{eq-27}
\delta f_a=f_a-\bar{f}_a
\end{equation}
where $\bar{f}_a$ is the one-particle ``equilibrium'' distribution around that the fluctuations occur. The latter is either some equilibrium solution of the stationary kinetic equation (\ref{eq-24}) or some of its large-scale solutions with scales exceeding those of the fluctuations such that the average of the fluctuation taken over these scales $\overline{\Delta f}_a=0$ vanishes. Neglecting the collision term by putting $\vec{C}_a=0$ the causal collisionless kinetic equation for the fluctuations then becomes by subtracting the fluctuation averaged kinetic equation
\begin{eqnarray}\label{eq-28}
&&\frac{\partial\,\delta f_a}{\partial t}+\frac{\vec{p}}{m_a}\vec{\cdot}\nabla_{\vec{q}}\:\delta f_a-\sum_b\frac{n_be_ae_b}{4\pi\epsilon_0}\int d^3p'd^3q'\Bigg\{\times\nonumber\\[-1ex]
&&\times\bigg(\nabla_{\vec{q}}+\frac{\vec{p}'}{m_bc^2}\frac{\partial}{\partial t}\bigg)-\frac{1}{m_bc^2}\nabla_{\vec{q}} \vec{\wedge}\vec{p}'\Bigg\}\:\vec{\cdot}\:\frac{\partial}{\partial\vec{p}}\:\bigg(\times\\
&&\times\frac{\delta{f_a(\vec{x},t)}\ \overline{ f_b(\vec{x}',t^R)}+\overline{f_a(\vec{x},t)}\ \delta f_b(\vec{x}',t^R)-\overline{\delta f_a\delta f_b}}{|\vec{q}-\vec{q}'|}\bigg)\!\! = 0\nonumber
\end{eqnarray}
This equation contains the correlations of the fluctuations $\overline{\delta f_a\delta f_b}$ which in a linearized theory are neglected. 

Clearly the above equations resemble the well-known approach to plasma kinetic theory. It should, however, be pointed out that even when in linear theory dropping the collision term on the right in a Klimontovich-Vlasov approach, the retardation effect remains in the third  term on the left-hand side in Eq. (\ref{eq-24}) which is the lowest order electromagnetic field-charged particle interaction term. 

\section{Discussion}

\subsection{Remarks}
The one-particle kinetic equation Eq. (\ref{eq-24}) obtained here is fundamental to all electromagnetic plasma interactions. Since these are electromagnetic, the purely electrostatic approximation when applied must be justified separately. This is not easy because in a strictly electrostatic approach the field response is instantaneous which contradicts electrodynamics and relativity on which it is based. It can be held up, if the information transport would occur by electrostatic waves only but still requires some assumption about the brevity of time delay. This assumption is that the electrostatic fluctuations occur on a vastly longer time scale than the travel time of light from the remotest position of particles. Thus one restricts oneself to sufficiently small plasma volumina in which the information transport may occur without some remarkable delay. 

Under such conditions Klimontovich-Vlasov theory applies, and the complications introduced by reference to the retarded time can be neglected. On the other hand, in very large volumes like in cosmical and astrophysical applications transport of information is provided by radiation transport and becomes rather slow. Hence remote volumina will not respond immediately and even not in light-propagation time which can then be treated again in the simplified theory. 

However, the current investigation is necessary as a clarification of two points: Firstly, that the interaction among different volumina in plasma in principle cannot be considered to occur instantaneously. Secondly, the inclusion of retarded times gives a clue for the direction of time -- as briefly discussed below -- which in many-particle-systems has only one direction, forward. Events are delayed by information transport and thus decorrelate even though they become relativistically synchronized by accounting for the information transport. This should necessarily contribute to dissipation because information becomes diffused by passing across the plasma from one particle to another.

\subsection{Direction of time}   
Reference to the retarded potentials and the effect of emission and absorption implies that already on the microscopic level there is a distinction between advanced and retarded effects. The delayed and integrated response of the charge and current densities at location $\vec{q}$ and time $t$ to the variation of the corresponding densities at all locations $\vec{q}'$ and $t'$ takes account of causality and thus of the direction of time. Ignoring the effect of time retardation the original Liouville equation is clearly symmetric in time. It does not distinguish between processes proceeding forward and backward in time. This is one of the big badly understood problems in physics which possibly resolves only on a macroscopic level. Microscopic equations seem, almost without exception, to be time-symmetric. However, when making reference to signal retardation in absorber theory this symmetry might be broken as suggested from the retarded time Eq. (\ref{eq-8}). By replacing $t\to-t$ one has
\begin{equation}
t'=-\big(t+|\vec{q}-\vec{q}'|/c\big)
\end{equation}
and thus, with constant velocity of light $c$, the negative retarded time $t'\to-t'$ becomes advanced thus breaking time symmetry in absorber theory. In order to restore retardation as required by the Wheeler-Feynman absorber theory one needs to redefine the velocity of light as $c\to-c$. Thus in a time-symmetric many-particle theory the negative time direction would come into accord with absorber theory only under the requirement that time velocity $c$ is negative there, i.e. one has to take the negative root $c=-1/\sqrt{\epsilon_0\mu_0}$. There is no obvious reason why this should be imposed thus becoming a philosophical question. Should $c$ be considered the inverse positive or negative root of the product of susceptibilities of the vacuum, or should $c$ be interpreted as a positive speed, the speed of light, with reference to a distance travelled by time in either positive or negative time? 

This question cannot be answered a priori. Absorber theory is restored in the second case in the causal many-particle theory. When considering the vacuum as a medium in which the dispersion of electromagnetic waves is described by a dispersion relation $\omega^2=k^2/\epsilon_0\mu_0$, interpreting this as the relation between photon energy and momentum, one has $\hbar\omega=\pm \hbar k/\sqrt{\mu_0\epsilon_0}$. Since photon energies should be real and positive, a negative sign of the root implies  negative wavenumbers/negative photon momenta. So it would require that time inversion also implies spatial inversion, which is not required by Maxwell's equations. One is thus tempted to accept that reference to retardation implies breaking time invariance in the Liouville equation and thus indicating that retardation on the classical microscopic level of kinetic theory of many particle systems may refer to the existence of an arrow of time which is directed forward in time.

\subsection{Summary}
The present investigation which extends Klimontovich's approach to kinetic plasma theory applies to systems of indistinguishable charged particles interacting via their self-consistent electromagnetic fields. One can trivially extend it to the presence of external fields like stationary or variable magnetic fields caused by external sources. The same procedure can also be applied to other classical fields since in all interactions transport of information from the agent to the absorber must take time. This is for instance the case in gases where sound waves or gravity waves can be excited and these transport the information from one fluid element to another place to affect the dynamics of other elements. In those cases it is not the photons but phonons that transport energy and information. Application to those systems lies outside our intention in the present work.   

%\begin{figure}
%\centerline{\includegraphics[width=0.5\textwidth,clip=]{epl-CR-fig2.pdf} }
%\caption{ } \label{epl-CR-fig2}
%\end{figure}

\begin{acknowledgement}
This work was part of a Visiting Scientist Programme at the International Space Science Institute Bern in 2007. Interest of the ISSI Directorate is acknowledged, as and in particular is the friendly hospitality of the ISSI staff. Thanks are directed to the ISSI system administrator S. Saliba for technical support and to the librarians Andrea Fischer and Irmela Schweizer for access to the library and literature.  
\end{acknowledgement}

%\newpage


\begin{thebibliography}{0}

%\bibitem[Vasyliunas(1968)]{vasyliunas1968} {Vasyliunas V.}: {J. Geophys. Res.} {73}, {2839, doi: 10.1029/JA073i009p02839, {1968}.}

%\bibitem[Christon et al.(1988)]{christon1988} {Christon S. et al.}: {J. Geophys. Res.} {93}, {2562, doi: 10.1029/JA093iA04p02562, {1988}.}

%\bibitem[Christon et al.(1991)]{christon1990} {Christon S. et al.}: {J. Geophys. Res.} {96}, {1, doi: 10.1029/90JA01633, {1991}.}
%\newpage
%\bibitem[Treumann(1999)]{treumann1999} {Treumann R.A.}: {Phys. Scripta} {59}, {204, doi: 10.1238/Physica.Regular.059a00204, {1999}}

%\bibitem[Treumann and Baumjohann(2014)]{treumann2014} {Treumann R.A. and W. Baumjohann}: {Front. Physics} {2}, {49, doi: 	
%10.3389/fphy.2014.00049, {2014}}








%\bibitem[Beatty et al.(2016)]{beatty2016} {Beatty J.J., Matthews J. and Wakely S.P.}: {Phys. Part. Phys., Chap. 29.} {Cosmic Rays}, {1}, {2016}


%\bibitem[Antoni et al.(2005)]{antoni2005} {Antoni T. et al. (Kaskade Collab.)}: {Astropart. Phys.} {10}, {1, doi:10.1016/j.astropartphys.2005.04.001,} {2005} 

%\bibitem[Amenomori et al.(2008)]{amenomori2008} {Amenomori M. et al.}: {Astrophys. J.} {687}, {1165, doi: 10.1086/529514, {2008}}

%\bibitem[Apel et al.(2011)]{apel2011} {Apel W.D. et al.}: {Phys. Rev. Lett.} {107}, {171104, doi: 10.1103/PhysRevLett.107.171104, {2011}}

%\bibitem[Aartsen et al.(2013)]{aartson2013} {Aartsen M.G. et al. (IceCube Collab.)}: {arXiv:1307.3795v1}, {2013}

%\bibitem[Abbasi et al.(2008)]{abbasi2008} {Abbasi R. et al. (HiRes Collab.)}: {Phys. Rev. Lett.} {100}, {101101, doi: 10.1103/PhysRevLett.100.101101}, {2008}

%\bibitem[Abraham et al.(2008)]{abraham2008} {Abraham J. et al. (Auger Collab.)}: {Phys. Rev. Lett.} {101}, {061101, doi: 10.1103/PhysRevLett.100.211101}, {2008}


%\bibitem[Ivanov et al.(2015)]{ivanov2015} {Ivanov D. et al. (Telescope Array Collab.)}: {Proc. Science} {ICRC}, {349}, {2015}


\bibitem[Jefimenko(1966)]{jefimenko1966} {Jefimenko O.D.}: {Electricity and Magnetism: An Introduction to the Theory of Electric and Magnetic Fields} (Appleton-Century-Crofts, New York, 1966).

%\bibitem[Kolmogorov(1991)]{kolmogorov1941} {Kolmogorov A.N.}: {Proc. Royal Soc. London A} {434}, {9 and 15, doi: 10.1098/rspa.1991.0076}, {1991}

\bibitem[Klimontovich(1967)]{klimontovich1967} {Klimontovich Y.L.}: {The Statistical Theory of Non-equilibrium Processes in a Plasma} (The MIT Press, Cambridge, MA, 1967).

\bibitem[Panofsky and Phillips(1962)]{panofsky1962} {Panofsky W.K.H. and Phillips M.}: {Classical Electricity and Magnetism} (Addison-Wesley, New York, 1962).



\bibitem[Wheeler and Feynman(1945)]{wheeler1945} {Wheeler J.A. and Feynman R.P.}: {Interaction with the absorber as the mechanism of radiation,} Rev. Mod. Phys. 17, 157-161, doi: 10.1103/RevModPhys.17.157., 1945.

\bibitem[Wheeler and Feynman(1949)]{wheeler1949} {Wheeler J.A. and Feynman R.P.}: {Classical electrodynamics in terms of direct interparticle action,} Rev. Mod. Phys. 21, 425-433, doi: 10.1103/RevModPhys.21.425., 1949.

%\bibitem[Frisch et al.(1978)]{frisch1978} {Frisch U.,  Sulem P.L. and Nelkin M.}: {J. Fluid Mech.} {87}, {719, doi: 10.1017/S0022112078001846}, {1978}

%\bibitem[Bykov and Treumann(2011)]{bykov2012} {Bykov A.M. and Treumann R.A.}: {Astron. Astrophys. Rev.} {19}, {42, doi: 10.1007/s00159-011-0042-8}, {2011}

\end{thebibliography}
\end{document}